\documentclass[10pt]{article}
\usepackage{a4wide}                      
\usepackage{epsf}                        
\usepackage{latexsym}                    
\usepackage{amsfonts}                    
\usepackage{amssymb}                     
\usepackage[spanish,activeacute]{babel} 

\newcommand{\sect}[1]{\setcounter{equation}{0}\section{#1}}

\def\be{\begin{equation}}
\def\ee{\end{equation}}
\def\bea{\begin{eqnarray}}
\def\eea{\end{eqnarray}}
\def\barray{\begin{array}}
\def\earray{\end{array}} 

\def\nnw{\nonumber \\ [.2cm]}
\def\vsp#1{\vspace{#1}}
\def\hsp#1{\hspace{#1}}
\def\!{!`}

\def\part{\partial}

\def\makeatletter{\catcode`\@=11}
\makeatletter
\def\mathbox#1{\hbox{$\m@th#1$}}%
\def\math@ccstyles#1#2#3#4#5#6#7{{\leavevmode
      \setbox0\mathbox{#6#7}%
      \setbox2\mathbox{#4#5}%
      \dimen@ #3%
      \baselineskip\z@\lineskiplimit#1\lineskip\z@
      \vbox{\ialign{##\crcr
             \hfil \kern #2\box2 \hfil\crcr
             \noalign{\kern\dimen@}%
             \hfil\box0\hfil\crcr}}}}
\def\mathaccstyles{\math@ccstyles\maxdimen}
\def\maththroughstyles{\math@ccstyles{-\maxdimen}}
\def\unity%
 {\maththroughstyles{.45\ht0}\z@\displaystyle {\mathchar"006C}\displaystyle 1}
\begin{document}

\centerline{\LARGE \bf La invariancia de la masa  }
\vspace{.4truecm}
\centerline{\LARGE \bf en Relatividad Especial}    
\vspace{0.7truecm}

\centerline{
    {\bf Bert Janssen }}

\vspace{.3cm}
\centerline{{\it Departamento de F\'isica Te\'orica y del Cosmos}}
\centerline{{\it Universidad de Granada}}
\centerline{{\it 18071 Granada}}
\centerline{Granada, noviembre 2012}

\vspace{.7truecm}

\centerline{\bf ABSTRACT}
\vspace{.3truecm}

\begin{center}
\begin{minipage}{13cm}
Se corrige un error que aparece con cierta frecuencia en la literatura divulgativa sobre la 
Teor\'ia de la Relatividad Especial: el hecho de que supuestamente la masa de un objeto en 
movimiento 
aumenta con la velocidad relativa entre el observador y el objeto. Explicaremos en este texto 
que es m\'as correcto afirmar que lo que aumenta es el momento y la energ\'ia del objeto, mientras 
que la masa es un invariante, independientemente del movimiento del observador como
del objeto mismo. 
\end{minipage}
\end{center}

\vspace{.2cm}
\section{Introducci\'on}

Con cierta frecuencia se puede encontrar en la literatura divulgativa sobre relatividad 
especial el concepto de {\it masa relativista} $m_r$ de un objeto, definida como
\be
m_r = \frac{m_0}{\sqrt{1 - \frac{v^2}{c^2}}},
\label{mrelativintro}
\ee
donde $m_0$ es la masa del objeto en resposo, $v$ es la velocidad relativa entre el objeto y 
el observador y $c$ es la velocidad de la luz. Originalmente esta f\'ormula fue introducida por 
Richard Tolman en 1912, quien cre\'ia que era la formula m\'as adecuada para la masa de un objeto en
movimiento. Conceptos relacionados fueron usados por Hendrik Anton Lorentz en su teor\'ia del 
electr\'on, donde introdujo la masa transversal y la masa longitudinal de un objeto por la manera 
en que aparec\'ian en las distintas componentes de la Segunda Ley de Newton,
\be
F_x  \,=  \,m_L \, a_x, \hsp{2cm}
F_y \, =  \,m_T \, a_y, \hsp{2cm}
F_z  \,=  \,m_T \, a_z,
\label{lorentz}
\ee  
donde 
\be
m_L = \frac{m_0}{\sqrt{1 - \frac{v^2}{c^2}}^3}, \hsp{2cm}
m_T = \frac{m_0}{\sqrt{1 - \frac{v^2}{c^2}}}.
\ee
Estas definiciones parec\'ian respaldadas por los experimentos de Kaufmann, Bucherer y Neumann entre 
1901 y 1905, que suger\'ian que la inercia de un objeto depende de su velocidad.

Sin embargo, el uso de los conceptos de masa relativista, masa longitudinal y masa transversal 
tienden a llevar a confusiones, dificultades e incluso errores, de modo que desde la mitad del 
Siglo XX se ha preferido trabajar con los conceptos de masa de reposo, energ\'ia y momento e
interpretar los resultados de Kaufmann, Bucherer y Neumann de manera diferente.  En 1948,
el mismo Albert Einstein escribi\'o en una carta al autor Lincoln Barnett, que en ese momento 
trabajaba en su libro divulgativo {\sl The Universe and Doctor Einstein}:
\begin{center}
\begin{minipage}{13cm}
{\it ``No es correcto introducir el concepto  de la masa $M = m/\sqrt{1 - v^2/c^2}$ de un cuerpo 
en movimiento, para el cual no se puede dar una definici\'on clara. Es mejor no introducir ning\'un
otro concepto de masa m\'as que la masa en reposo $m$. En lugar de introducir $M$ es mejor 
mencionar la expresi\'on para el momento y la energ\'ia de un objeto en movimiento.'' }
\end{minipage}
\end{center}

En este escrito demostraremos que la introducci\'on de la masa relativista surge de una 
formulaci\'on err\'onea de la Segunda Ley de Newton relativista y explicaremos los resultados de 
los experimentos de Kaufmann, Bucherer y Neumann en t\'erminos del momento y la energ\'ia 
relativista. En la secci\'on 2 introduciremos dos formulaciones en general inequivalentes de la 
Segunda Ley de Newton en la mec\'anica newtoniana y explicaremos cu\'al de las dos es m\'as correcta.
En la secci\'on 3 extenderemos esta formulaci\'on para que incluya tambi\'en efectos relativistas y 
explicaremos la interpretaci\'on moderna de los resultados de Kaufmann, Bucherer y Neumann. 
Finalmente en la secci\'on 4 explicaremos el error en la derivaci\'on de la Segunda Ley de Newton
en t\'erminos de la masa relativista y demostraremos que en realidad la masa en reposo tiene
un valor constante para cualquier observador inercial. En la secci\'on 5 discutiremos dos paradojas
generadas por el uso de la masa relativista y en la secci\'on 6 resumiremos las ºconclusiones.

\sect{La Segunda Ley de Newton en la mec\'anica newtoniana}

En la mec\'anica newtoniana se pueden encontrar dos expresiones distintas para la famosa Segunda Ley
de Newton, que relaciona las fuerzas que actuan sobre un objeto con los cambios de velocidad
de este objeto. La formulaci\'on m\'as frecuente es
\be
\vec F = m \vec a,
\label{F=ma}
\ee 
donde $m$ es la masa del objeto y $\vec a$ es la 
aceleraci\'on debida a la fuerza $\vec F$ que actua sobre esta. Matem\'aticamente hablando
la aceleraci\'on es la derivada de la velocidad con respecto al tiempo,
\be
\vec a (t) = \frac{d\vec v(t)}{dt} = \frac{d^2\vec r(t)}{dt^2},
\ee   
y parametriza c\'omo cambia la velocidad $\vec v(t)$ con el tiempo.   

La otra formulaci\'on de la Segunda Ley de Newton, menos frecuente, relaciona la fuerza actuando 
sobre un objeto con cambios del momento lineal:
\be
\vec F =\frac{d\vec p}{dt},
\label{F=dp/dt}
\ee
 donde $\vec p$ es el momento lineal,
\be
\vec p = m \vec v.
\label{p=mv}
\ee

Las dos formulaciones (\ref{F=ma}) y (\ref{F=dp/dt}) son equivalentes en el caso donde la masa 
del objeto es constante, puesto que entonces
\be
\vec  F \ = \ \frac{d\vec p}{dt} \ = \ \frac{d(m\vec v)}{dt} 
        \ = \ m\frac{d\vec v}{dt}  \ = \ m\vec a.
\ee  
Sin embargo, en el caso donde la masa de objeto cambia con el tiempo, $m=m(t)$ (por ejemplo, 
un cohete que va expulsando gases al acelerar, o un carro que va acumulando lluvia mientras que 
se mueve), 
las dos f\'ormulas (\ref{F=ma}) y (\ref{F=dp/dt}) no son equivalentes. De (\ref{F=dp/dt}) 
tendr\'iamos que 
\be
\vec  F  \ = \ \frac{d(m(t)\vec v)}{dt} 
        \ = \ m(t) \vec a \ + \  \frac{dm(t)}{dt} \vec v. 
\ee
En otras palabras, (\ref{F=ma}) es claramente un caso especial de (\ref{F=dp/dt}), cuando la
masa es constante, y la f\'ormula correcta es, en toda su generalidad, 
(\ref{F=dp/dt}).\footnote{La raz\'on por qu\'e se encuentra m\'as la formulaci\'on (\ref{F=ma}) es 
    porque en la gran mayor\'ia de los casos tratados, la masa de un objeto es constante, en cuyo 
    caso esta formulaci\'on m\'as intuitiva, efectivamente, se aplica.}

Se puede demostrar esto con el experimento sencillo de un vag\'on de tren que se desliza sin 
rozamiento por
un carril horizontal. En ausencias de fuerzas externas, (\ref{F=dp/dt}) implica que el momento 
lineal $\vec p = m \vec v$ es constante a lo largo de movimiento. Si la masa del vag\'on tambi\'en 
es constante, el vag\'on se mueve de manera uniforme rectilinea, con velocidad constante. Sin 
embargo cuando empieza a llover, el vag\'on empezar\'a a acumular agua y el aumento de 
masa repercutir\'a en el movimiento del vag\'on. Concretamente, si la lluvia cae perfectamente 
vertical, no ejercer\'a ninguna fuerza en la direcci\'on de movimiento, de modo que el momento
$\vec p$ a\'un se conserva. Sin embargo, al aumentar la masa del vag\'on, va a disminuir la 
velocidad $\vec v$ de tal forma que el producto $\vec p = m \vec v$ se mantenga constante. Este
fen\'omeno se ha observado en numerosos casos y es un efecto importante en ingenier\'ia 
aeroespacial.

\sect{La Segunda Ley de Newton en Relatividad Especial }

La teor\'ia de la relatividad especial modifica y generaliza la mec\'anica newtoniana a 
velocidades cercanas a la velocidad de luz, $v \sim c$. Concretamente afirma que la velocidad
de la luz es la misma para todos los observadores inerciales y que adem\'as es la velocidad
m\'axima con que se puede mover cualquier observador, objeto o incluso informaci\'on.

Una consecuencia directa de esta \'ultima afirmaci\'on es que la formulaci\'on (\ref{F=ma}) no puede 
ser cierta, puesto que implicar\'ia que una fuerza constante podri\'a acelerar una masa hasta 
velocidades arbitrariamente grandes, mucho m\'as que la velocidad de la luz. Otra vez la 
formulaci\'on correcta de la Segunda Ley de Newton es el equivalente relativista de (\ref{F=dp/dt}),
\be
\vec F =\frac{d\vec p}{dt}.
\label{F=dp/dt2}
\ee
La gran diferencia entre la f\'ormula relativista (\ref{F=dp/dt2}) y la newtoniana  (\ref{F=dp/dt}) 
es la definici\'on de momento $\vec p$. Concretamente el momento relativista est\'a definido como
\be
\vec p = \frac{m_0 \vec v}{\sqrt{1 - \frac{v^2}{c^2}}},
\label{prelativ}
\ee
donde $m_0$ es la masa del objeto, medida en el sistema que est\'a en reposo con respecto a objeto
y $\vec v$ es la velocidad relativa entre el objeto y el observador.  
Obs\'ervese que si el objeto se mueve con velocidades mucho m\'as peque\~nas que la velocidad de la 
luz, $v \ll c$, el factor en el denominador de (\ref{prelativ}) es pr\'acticamente 1 y en este 
l\'imite la diferencia entre (\ref{p=mv}) y (\ref{prelativ}) es despreciable. Sin embargo, cuando
el objeto se mueve con una velocidad cercana a la velocidad de la luz, la presencia del factor 
en el denominador causar\'a grandes correcciones a la expresi\'on (\ref{prelativ}), de modo que 
el momento relativista de un objeto movi\'endose a velocidades relativistas es mucho mayor que 
el momento newtoniano. 

N\'otese tambi\'en que las expresiones (\ref{F=dp/dt2}) y (\ref{prelativ}) implican que ning\'un 
objeto puede alcanzar, ni sobrepasar la velocidad de la luz: al aplicar una fuerza $\vec F$ sobre 
un objeto resulta en un cambio del momento relativista $\vec p$ (m\'as que en un cambio de la 
velocidad $\vec v$). Dado que la mayor contribuci\'on al momento relativista de un objeto que 
se mueve con 
velocidades cercanas a $c$ viene del  denominador de  (\ref{prelativ}), cualquier fuerza finita 
aumentar\'a sobre todo el factor $1/\sqrt{1 - \frac{v^2}{c^2}}$, m\'as que el factor $m_0\vec  v$. 
Adem\'as este efecto es mayor, cuando m\'as que se acerca la velocidad del objeto a la de la luz.
Para acelerar el objeto hasta la velocidad de la luz, hace falta una fuerza infinitamente grande, 
la cual no existe en la naturaleza. En la vieja interpretaci\'on de Tolman y Kaufmann, Bucherer y 
Neumann se cre\'ia que esto implicaba que la inercia del objeto crece en funci\'on de su velocidad, 
pero ahora vemos que es mucho m\'as natural decir que lo que aumenta es el momento relativista  
(\ref{prelativ}).

\sect{La constancia de la masa en Relatividad Especial }

Con cierta frecuencia se puede encontrar que algunos libros sobre relatividad especial intentan
escribir la Segunda de Ley de Newton relativista  (\ref{F=dp/dt2}) en una forma m\'as cercana a
 (\ref{F=ma}). Para ello suelen definir la masa relativista de un objeto como
\be
m_r = \frac{m_0}{\sqrt{1 - \frac{v^2}{c^2}}},
\label{mrelativ}
\ee
y luego escribir (\ref{F=dp/dt2}) como
\be
\vec F = \frac{d}{dt}\left(\frac{m_0 \vec v}{\sqrt{1 - \frac{v^2}{c^2}}}\right) 
       = \frac{m_0 \vec a}{\sqrt{1 - \frac{v^2}{c^2}}}
       = m_r  \vec a
\hsp{1cm}
\mbox{\bf\!\!F\'ormula \ \ err\'onea!!}
 \label{error}
\ee
Sin embargo esta derivaci\'on es err\'onea, por la misma raz\'on que (\ref{F=ma}) y (\ref{F=dp/dt})
en general no son equivalentes. Concretamente, el error de (\ref{error}) est\'a en la segunda 
igualdad: la derivada $d/dt$ no solo act\'ua sobre la velocidad que 
aparece en el numerador, sino tambi\'en en el factor $v^2/c^2$ en el denominador. Eso es 
precisamente lo que explicamos al final de la secci\'on anterior: que una fuerza resulta apenas
en un aumento de velocidad, pero aumenta much\'isimo el momento relativista.  
La derivaci\'on correcta de (\ref{error}) ser\'ia
\be
\vec F \ = \ \frac{d}{dt}\left(\frac{m_0 \vec v}{\sqrt{1 - \frac{v^2}{c^2}}}\right) 
       \ = \  \frac{m_0(\vec v \cdot \vec a) \vec v}{{\sqrt{1 - \frac{v^2}{c^2}}^3}} 
                  \ + \ \frac{m_0\vec a}{\sqrt{1 - \frac{v^2}{c^2}}}.
\ee
Observese que es por lo tanto el primer t\'ermino el que impide escribir la Segunda Ley de Newton
como  (\ref{error}), lo que hace la definici\'on (\ref{mrelativ}) innecesaria, e incluso confusa. 

Finalmente hay otra manera de ver que la masa $m_0$ en relatividad especial es una cantidad
constante, independiente del estado de movimiento del observador y del objeto. De la misma manera 
que hemos  definido el momento relativista en (\ref{prelativ}) podemos definir la energ\'ia 
relativista como  
\be
E = \frac{m_0c^2}{\sqrt{1 - \frac{v^2}{c^2}}}.
\label{Erelativ}
\ee
Para velocidades mucho m\'as bajas que la velocidad de la luz, esta expresi\'on se simplifica 
(a trav\'es de un desarrollo de Taylor en $\frac{v^2}{c^2}$) a 
\be
E \approx m_0 c^2 + \frac{1}{2} m_0 v^2 + ...
\ee 
En otras palabras, a velocidades cotidianas recuperamos la expresi\'on netwoniana para la energ\'ia 
cin\'etica, m\'as el famoso t\'ermino para la energ\'ia de reposo, $E=m_0 c^2$. Sin embargo, lo m\'as
interesante es que las expresiones  (\ref{prelativ}) y (\ref{Erelativ}) combinan en la relaci\'on
\bea
E^2 - p_x^2 c^2  - p_y^2 c^2 - p_z^2 c^2 &=&
\frac{m^2_0c^4}{1 - \frac{v^2}{c^2}} \ - \ \frac{m^2_0v_x^2c^2}{1 - \frac{v^2}{c^2}}  
   \ - \ \frac{m^2_0v_y^2c^2}{1 - \frac{v^2}{c^2}}  \ - \ \frac{m^2_0v_z^2c^2}{1 - \frac{v^2}{c^2}}  
                 \nnw
&=& m_0^2 c^4 \ \frac{1 - (v_x^2 + v_y^2 + v_z)/c^2}{1 - \frac{v^2}{c^2}} \nnw
&=& m_0^2 c^4.
\label{E^2-p^2}
\eea 
En otras palabras, un observador puede determinar la masa de un objeto si conoce su energ\'ia $E$
y su momento $\vec p$ relativista.

Ahora, tanto la energ\'ia (\ref{Erelativ}) como el momento  (\ref{prelativ}) de un objeto dependen 
de la velocidad $v$ con que se mueven con respecto a un observador. Un observador ${\cal O}'$ que 
est\'a en reposo respecto a la part\'icula ver\'a una energ\'ia 
$E'=m_0 c^2$ y un momento $\vec p \ {}' = 0$, mientras que el observador ${\cal O}$ ver\'a un 
momento no-nulo y una energ\'ia m\'as grande. Resulta que $E$ y $\vec p$ est\'an relacionados con 
 $E'$ y $\vec p \ {}'$ a trav\'es de una transformaci\'on de Lorentz
\bea
E'= \frac{E - v p_x}{\sqrt{1-v^2/c^2}}, \hsp {1cm}
p_x'= \frac{p_x-vE/c^2}{\sqrt{1-v^2/c^2}}, \hsp {1cm}
p'_y = p_y, \hsp {1cm}
p'_z = p_z.
\label{transfE}
\eea

Obs\'ervese ahora que  la relaci\'on (\ref{E^2-p^2}) es invariante bajo las transformaciones 
(\ref{transfE}): el observador  $\cal O'$, que mide energ\'ia $E'$ y momento $\vec p \ {}'$ puede 
traducir sus resultados en terminos del $E$ y $\vec p$ medidos por  $\cal O$ como
 \bea
&& (E')^2 - (p'_x c)^2  - (p'_y c)^2 - (p'_z c)^2 \ = \ \nnw
&& \hsp{2cm}
   = \left(\frac{E - v p_x}{\sqrt{1-v^2/c^2}} \right)^2 
           \ -  \  \left(\frac{p_x-vE/c^2}{\sqrt{1-v^2/c^2}}\right)^2c^2
           - p^2_y c^2 - p^2_z c^2  \nnw
 && \hsp{2cm} = \frac{E^2 - 2vEp_x + v^2 p_x^2}{1-v^2/c^2} 
               \ - \ \frac{p_x^2c^2 - 2vEp_x + v^2 E^2/c^2}{1-v^2/c^2} - p^2_y c^2 - p^2_z c^2  \nnw
 && \hsp{2cm} = \frac{(1-v^2/c^2) (E^2 - p^2_x c^2)}{1-v^2/c^2} - p^2_y c^2 - p^2_z c^2  \nnw
&& \hsp{2cm} = E^2 - p^2_x c^2  - p^2_y c^2 - p^2_z c^2  \nnw
 && \hsp{2cm} = m_0^2 c^4.
\label{E^2-p^2bis}
\eea 
En otras palabras, un observador $\cal O$, que mide una energ\'ia $E$ y un momento $\vec p$, y un 
obseravdor $\cal O'$, que mide una energ\'ia $E'$ y un momento $\vec p \ {}'$, obtendr\'an el mismo 
valor para la masa $m_0$ del objeto:
\be
E^2 - p_x^2 c^2  - p_y^2 c^2 - p_z^2 c^2 \ = \ m_0^2 c^4 \ = \
  (E')^2 - (p'_x c)^2  - (p'_y c)^2 - (p'_z c)^2. 
\ee
La masa $m_0$ de un objeto es por lo tanto un invariante, 
independiente del estado de movimiento del observador y del objeto mismo, puesto que 
su valor es el mismo cualquier conjunto de observadores conectados 
a trav\'es de una transformaci\'on de Lorentz. 

\sect{Paradojas}

Como ejemplo de que el concepto de masa relativista puede llegar a conclusiones err\'oneas, 
presentamos algunas paradojas que surgen al asumir que la masa aumenta como funci\'on de la 
velocidad relativa entre observador y objeto. En tal caso, la conclusi\'on de que un objeto 
en movimiento tiende a formar un agujero negro parece casi inevitable: el aumento de masa, 
combinado con la dismunici\'on de volumen debido a la contracci\'on de Lorentz sugiere que 
cierto momento hay una gran cantidad de masa comprimida dentro de un radio menor que radio 
de Schwarzschild, por lo que se formar\'ia un agujero negro.

Sin embargo, esta conclusi\'on es err\'onea, b\'asicamente por el hecho de que el efecto de aumento
de masa relativista no existe, como hemos explicado arriba. La falsedad de esta conclusi\'on se 
puede ver en la facilidad con que se pueden crear paradojas y contradicciones, si se asume lo 
contrario. Estas paradojas usan el Principio de la Relatividad, que afirma que cada uno de dos 
observadores en relativo movimiento uniforme puede considerarse a s\'i mismo en reposo y el otro
movi\'endose. Por lo tanto, las paradojas surgen 
debido al hecho de que el supuesto agujero negro solo se manifestar\'ia para el observador que se
mueve con respecto a la masa considerada, mientras que el observador en reposo con respecto a 
esta masa no notar\'ia nada particular, ya que no medir\'ia un aumento de masa relativista. 
Sin embargo los hechos f\'isicos, como los efectos que el supuesto agujero negro tendr\'ia sobre su 
entorno, deber\'ian ser objetivos y no pueden depender del quien lo observa.\\ 

\noindent
{\bf La paradoja de los granitos de arena}

Supongamos que se colocan dos granitos de arena en el espacio vac\'io, a una distancia el 
uno del otro tan grande que la fuerza gravitatoria es completamente despreciable. De este modo
los dos granitos no notan su mutua presencia y si  inicialmente est\'an en reposo entre ellos y
con respecto a un observador $\cal O$, podemos suponer que seguir\'an en reposo entre ellos y con 
respecto a $\cal O$.

Un observador $\cal O'$ que se mueve con respecto a  $\cal O$ y a los granitos de arena con una 
velocidad arbitrariamente cercana a la velocidad de la luz, se considerar\'a a s\'i mismo en reposo
 y afirmar\'a que son los granitos y el observador  $\cal O$ los que se est\'an moviendo con 
velocidades
relativistas. Si la teor\'ia del aumento de la masa y la formaci\'on de agujeros negros fuera 
correcta, el observador ver\'ia los granitos como dos agujeros negros con masas arbitrariamente 
grandes a una distancia arbitrariamente peque\~na entre ellos, ya que la distancia relativa sufre
una contracci\'on de Lorentz. Los dos agujeros negros experimentar\'ian una atracci\'on gravitatoria 
enorme entre ellos y colapsar\'ian en seguida, formando un solo agujero negro.  

Por lo tanto, los dos observadores $\cal O$ y $\cal O'$ ver\'ian dos procesos f\'isicos distintos
y mutuamente incompatibles: desde un punto de vista las masas colapsan, desde otro punto de vista
las masas siguen en reposo. Sin embargo la realidad es \'unica: o colapsan o no colapsan, lo que 
nos lleva a una paradoja que solo se puede resolver desestimando el aumento de masa realtivista. 
\\

\noindent
{\bf La paradoja de las hormigas comunicadoras}

Imaginamos una esfera (un globo terraqueo) maciza, de 3 metros de di\'ametro, en la que se 
colocan dos hormigas, una en el polo norte y otro en el polo sur. Las dos hormigas se comunican 
constantemente mandando se\~nales entre ellas (por ejemplo dando golpes al globo de modo que las 
vibraciones se propaguen por el material, o simplemente por una fibra \'optica que se extiende entre
los polos a lo largo del meridiano de Greenwich).

Un observador externo, que ve pasar la esfera con velocidades relativistas, la ver\'ia 
Lorentz-contraido en la direcci\'on del movimiento (digamos el plano ecuatorial en la direcci\'on del
meridiano de Greenwich), pero no en las direcciones transversales, ya que la contracci\'on de Lorentz
es puramente longitudinal. Por lo tanto, para este observador, el globo tiene la forma de un 
elipsoide, pero la distancia entre las hormigas sigue siendo 3 metros.  

Si la teor\'ia del aumento de masa y la formaci\'on de agujeros negros fuera correcta, 
podr\'iamos imaginarnos que la esfera se moviera con la velocidad adecuada para que se formara un 
agujero negro con un horizonte con un tama\~no m\'as peque\~no que el tama\~no transversal de la 
esfera, digamos 1,5 metros, de modo que la zona cercana al ecuador estar\'ia dentro del horizonte,
pero los polos no (recuerda que la contracci\'on de Lorentz solo es longitudinal). En este caso
el observador ver\'ia que las se\~nales que intercambian las dos hormigas entran en el agujero negro
cuando se acercan al ecuador, pero tambi\'en salen de \'el cuando se acercan al otro polo, lo que 
contradice la propia definici\'on de agujero negro.

\sect{Conclusiones}

El concepto de masa relativista, definida en (\ref{mrelativ}) a base de una derivaci\'on 
err\'onea de la Segunda Ley de Newton relativista,  es un concepto innecesario y 
poco pr\'actico, cuyo uso suele llevar a confusi\'on y errores conceptuales. Es conceptualmente 
m\'as correcto y m\'as claro explicar la din\'amica relativista  en t\'erminos de la masa en reposo 
$m_0$, la energ\'ia y el momento relativista, $E$ y $\vec p$, definidos respectivamente 
en (\ref{Erelativ}) y (\ref{prelativ}).  De esta manera, se trabaja con cantidades f\'isicas que
transforman de una manera bien definida bajo transformaciones de Lorentz, que relacionan 
distintos observadores inerciales. Concretamente, un c\'alculo sencillo demuesta que la masa
en reposo de un objeto es un invariante, que tiene el mismo valor para todos los observadores, 
independientemente de su estado de movimiento o el del objeto mismo. \\

 
\newpage
\noindent
{\Large{\bf Bibliograf\'ia}}
\begin{enumerate}
\item B. Janssen, {\sl Teor\'ia de la Relatividad General}, Universidad de Granada (2013), \\
  {\tt http://www.ugr.es/local/bjanssen/text/BertJanssen-RelatividadGeneral.pdf} 

\item B. Janssen, {\sl Breve repaso de la Relatividad Especial}, Universidad de Granada (2005), \\
  {\tt  http://www.ugr.es/local/bjanssen/text/repaso.pdf}\\
  DOI: 10.13140/RG.2.2.27449.52328.
  
\item L. Okun, {\sl The concept of mass}, Physics Today, June 1989.
\item J. Solbes, F.J. Botella, H. P\'erez, F. Tar\'in, {\sl Algunas consideraciones sobre la masa},
      Revista Espa\~nola de F\'isica {\bf 16} (1), 2002.
\item Wikipedia: \\
  {\tt http://en.wikipedia.org/wiki/Mass\_in\_special\_relativity}\\
  {\tt http://en.wikipedia.org/wiki/Kaufmann-Bucherer-Neumann\_experiments}

\end{enumerate}

\end{document}